\documentstyle[aps,prb,epsf,amsfonts,amssymb,preprint]{revtex}

\begin{document}
\draft

\title
{
\LARGE \bf
Hall tunneling of vortices in superclean superconductors}
\author{{\bf ${ \rm \bf D. A. Gorokhov}^{*}$  and  G. Blatter}
\footnote{{\it Theoretische Physik, ETH-H\"onggerberg,
CH-8093 Z\"urich, Switzerland}; 
e-mail: gorokhov@itp.phys.ethz.ch} 
}
\maketitle
\begin{abstract}
We discuss the main features of Hall tunneling
of pancake vortices in superclean high-$T_{c}$ superconductors.
The general formalism for the calculation of the lifetime
of a vortex pinned in a metastable configuration is described.
The results are applied to the problem of quantum tunneling
of a pancake vortex from a columnar defect in the limit
of a small driving current.

\end{abstract}
\vskip1.5cm

An external magnetic field  
penetrates a type-II superconductor in the form
of vortices --- flux tubes enclosing a  quantum 
$\Phi_{0}={\pi\hbar c}/{e}$
of magnetic flux.
The new class of high temperature superconductors
exhibits a large variety of vortex related phenomena, which can be tuned
by specific material parameters\cite{Blatter}, 
e.g., the material anisotropy and
the macroscopic and microscopic amount of disorder,
the former determining vortex pinning, whereas the latter is responsible
for the vortex dynamics. Here we are interested in highly anisotropic (layered)
clean material. A single vortex then can be represented as a chain
of pancake vortices. 
The (low frequency) vortex
equation of motion takes the form
\begin{equation}
\frac{\Phi_{0}}{c}{\bf j}\wedge{\bf n}
-{\bf \nabla}U_{pin}
=\eta {\bf v}+m{\bf \dot v}+
\alpha{\bf v}\wedge {\bf n}, 
\label{dwizhenie}
\end{equation}
where the dynamical forces (the Hall term $\alpha{\bf v}\wedge {\bf n}$,
the massive term ($m{\bf \dot v}$), and the dissipative term ($-\eta {\bf v}$))
are balanced by the Lorentz and pinning forces.
In this paper we consider pancake vortices parallel to the 
$c$-axis of a layered superconductor. 

Usually one can neglect
the massive term. The ratio of the 
dynamical coefficients $\alpha$ and $\eta$ 
can be estimated as 
${\alpha}/{\eta}\simeq ({\Delta}/{{\epsilon}_{F}})(l/\xi ),$
with $\Delta$ the gap, $\epsilon_{F}$ the Fermi energy,
$l$ the electron mean free path, and $\xi$ the ($ab$-plane) coherence length.
Typically, in high-$T_{c}$ superconductors 
${\Delta}/{{\epsilon}_{F}}\simeq {1}/{10}$--$1/{20}$,
$l\simeq {\rm 700~\AA}$, and $\xi\simeq {\rm 20~\AA}$, i.e., we obtain
${\alpha}/{\eta}\simeq 1$; in conventional superconductors
${\alpha}/{\eta}\ll 1$ due to the small ratio ${\Delta}/{\epsilon_{F}}$,
as typically $\Delta\sim 10$ K and $\epsilon_{F}\sim 10^{4}$ K).
Indeed, recent experiments suggest that the regime ${\alpha}/{\eta}\agt 1$
can be realized. In this situation, we can neglect the dissipative
term in the vortex equation of motion (\ref{dwizhenie}) and
describe the dynamics of the pancake vortex
with the Hall term alone. As a result, the problem becomes equivalent
to  that of a particle moving in a strong magnetic field.

The Lagrangian describing a pancake vortex in the presence
of a potential $U(x,y)$ produced
by pinning and Lorentz forces can be written in the form
\begin{equation}
L=\alpha {\dot x}y - U(x,y).
\label{lagr}
\end{equation}
and the associated equations 
of motion read
\begin{equation}
{\alpha}{\dot x}=\frac{\partial U(x,y)}{\partial y},
\label{1}
\end{equation}
\begin{equation}
{\alpha}{\dot y}=-\frac{\partial U(x,y)}{\partial x}.
\label{2}
\end{equation}
These equations of motion are equivalent to those
produced by the 1D Hamiltonian $H(x,p)=U(x,{p}/{\alpha})$,
i.e., our problem is effectively one dimensional\cite{Volovik}.
We consider a pinning potential of the form
$U_{pin}=U\left (\sqrt{x^{2}+y^{2}}\right )$ producing
a stable trapped vortex state
(such pinning centers can be 
created by heavy ion irradiation).
 For a nonzero external driving current ${\bf j}$ 
this state becomes metastable, resulting in a finite lifetime
for the trapped pancake vortex. At low temperatures only tunneling
from the ground state is relevant, whereas at higher temperatures 
thermal effects become important. In this note we study the
vortex depinning at arbitrary temperatures. The problem 
of Hall tunneling of
vortices in high-$T_{c}$ superconductors has been studied 
for various geometries and for different external currents,
see Refs.~3-9. Below we follow the analysis of Ref.~9 (see also Ref.~10).

The (real time) Lagrangian (\ref{lagr}) produces the 
Euclidean action (we carry out the substitution 
$S=\int L\thinspace dt\rightarrow -iS$ and $t\rightarrow -i\tau$)
\begin{equation}
S_{\rm Eucl}=\int\limits_{-{\hbar}/{2T}}^{+{\hbar}/{2T}}
\left [-i\alpha{\dot x}y +U_{0}\left (\sqrt{x^{2}+y^{2}}\right )
\right ]\thinspace d\tau.
\end{equation}
In general, the imaginary unit appearing in the Euclidean action
renders the saddle-point solution complex.
However,  in the present case we can 
reduce the complex
problem to a real one\cite{Blatter} via the additional
transformation $y\rightarrow iy$,
replacing $U_{0}\left (\sqrt{x^{2}+y^{2}}\right )$
by $U_{0}\left (\sqrt{x^{2}-y^{2}}\right )$.
After this transformation the Lagrangian
exhibits a saddle-point solution and we can calculate the
decay rate following the standard technique. 
The effective tunneling problem 
 is described by the Euclidean action
\begin{equation}
S_{\rm Eucl}[x(\tau ),y(\tau )]=\int_{-{\hbar}/{2T}}^{+{\hbar}/{2T}}
\left [
\alpha{\dot x}y+U_{0}\right (\sqrt{x^{2}-y^{2}}\left )-Fx\right ]d\tau.
\label{ea}
\end{equation}

\centerline{\epsfxsize=4.8cm \epsfbox{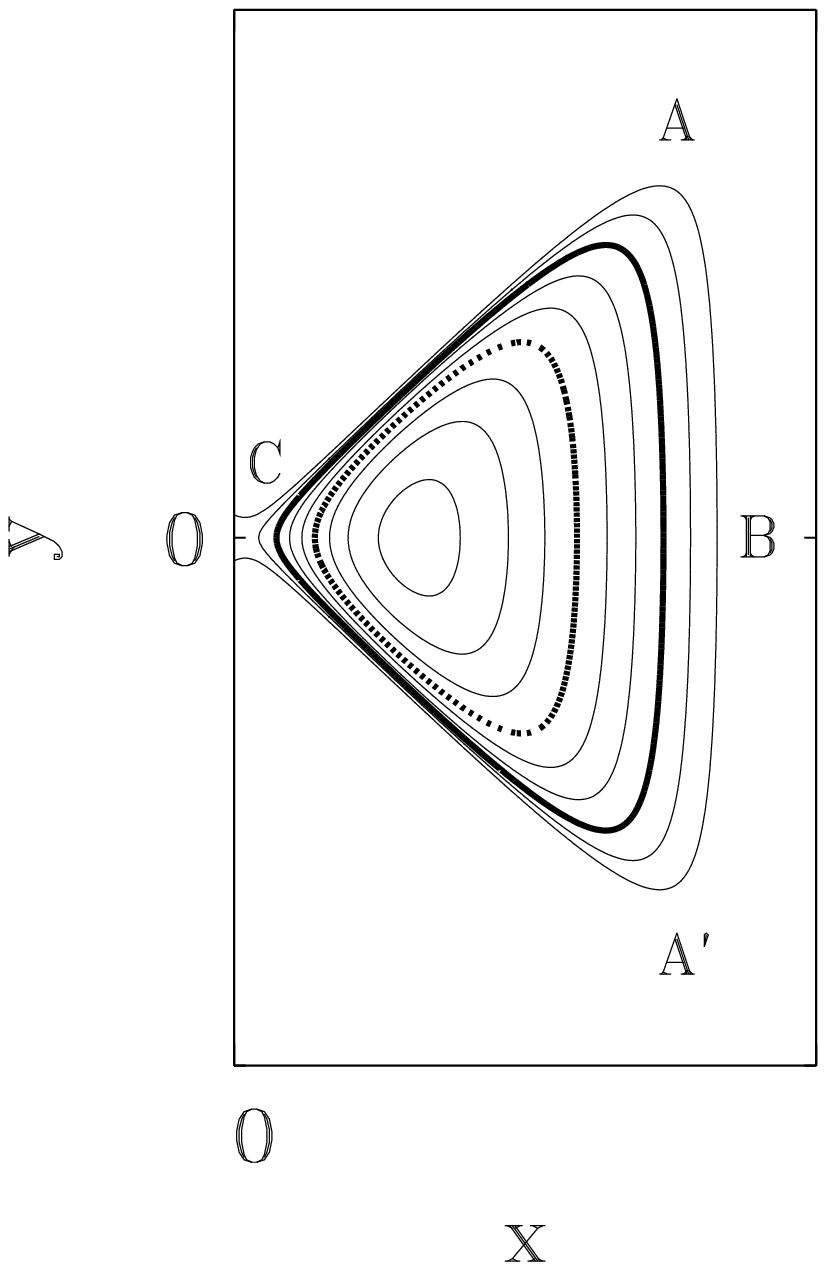}}
{\vskip0.5cm\footnotesize {\bf Fig.1}~Quasiclassical $(x,y)$-trajectories corresponding to 
the tunneling of the vortex. The Euclidean action corresponding
to the zero temperature trajectory 
(thick line)
is equal to the encircled area.
 The thick dotted line  marks a finite temperature
bounce trajectory..}
\vskip0.3cm
\hskip-0.7cm

The bounce trajectories are given by the condition
$U_{0}\left (\sqrt{x^{2}-y^{2}}\right )-Fx={\rm Const}$
and are shown in Fig.~1. 
The problem studied here exhibits the interesting
feature of geometric quantization: the action is
proportional to the area encircled by the pancake vortex
during its motion
in imaginary time. At zero temperature  
$U_{0}\left (\sqrt{x^{2}-y^{2}}\right )-Fx=0$
and $S_{\rm Eucl}=\int_{-\infty}^{+\infty}\alpha{\dot x}y\thinspace d\tau =
\alpha\oint x\thinspace dy$.
The Euclidean action at finite temperatures can be calculated 
using the following technique: The imaginary time 
trajectory at a finite energy $E$ is given by the equation
$U_{0}\left (\sqrt{x^{2}-y^{2}}\right )-Fx=E$. The trajectories
are distributed with the Boltzmann weight, i.e, we have
to find the trajectory for which 
$-{S(E)}/{\hbar}-{E}/{T}$ takes a maximal value.
 In the limit of small external currents the problem can be solved
analytically in the whole temperature range (see Ref.~9).
 The Euclidean action
(which determines the decay rate $\Gamma$ to exponential accuracy,
$\Gamma\sim\exp\left (-{S_{\rm Eucl}}/{\hbar}\right )$)  
can be written in the form

\begin{equation}
S_{\rm Eucl}=
\left\{ \begin{array}{r@{\quad\quad}l} 
\displaystyle{\alpha{\left (\frac{U_{0}}{F}\right )}^{2},} 
& {T<\frac{\hbar F^{2}}{2\alpha U_{0}}\equiv T_{1},}\\ \noalign{\vskip 5 pt}
\displaystyle{\frac{\hbar U_{0}}{T}-{\frac{\hbar^{2}F^{2}}{4\alpha T^{2}}}}, & 
{T_{1}<T,}\end{array}\right.
\label{otwet}
\end{equation}
where $U_{0}$ is the depth of the pinning potential and 
$F$ denotes the external force
acting on the pancake vortex. The function $S_{\rm Eucl}(T)$
is a constant below the temperature $T_{1}$,
whereas above $T_{1}$ thermal corrections become relevant.
We investigated in detail the crossover from the thermal
assisted quantum regime to the region of purely thermal activation.
For the pinning potential exhibiting a large-distance
asymptotic behavior of the form $U(r)\rightarrow U_{0}-B/{r^{2}}$, 
the transition is second-order like with the crossover temperature
given by the formula 
\begin{equation}
T_{c}=\frac{\sqrt{3}}{2^{{4}/{3}}}
\frac{\hbar F^{{4}/{3}}}{\alpha B^{{1}/{3}}}.
\label{Tc}
\end{equation}
For $T>T_{c}$ 
the escape process is purely thermal with
$S_{\rm Eucl}= {\hbar U_{0}}/{T}$, i.e. the decay
is due to thermal activation alone.
The evolution of the Euclidean action with increasing temperature is
sketched in Fig.~2.

\vskip-1cm
\centerline{\epsfxsize=10cm \epsfbox{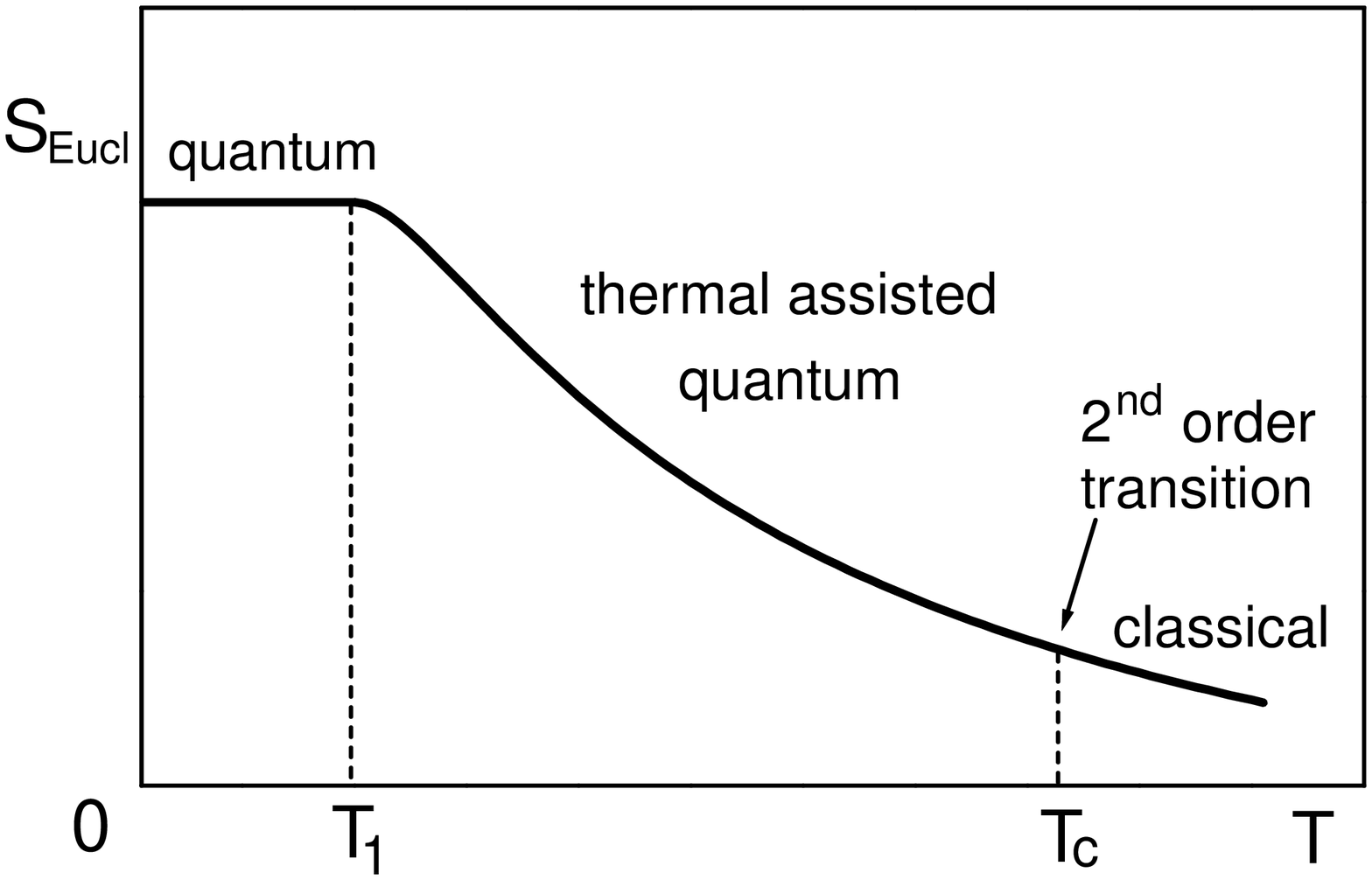}}
{\vskip0.5cm\footnotesize {\bf Fig.2}~Euclidean 
action as a function of temperature. At $T<{ T_{1}}=
{\hbar F^{2}}/{2\alpha U_{0}},$ $S_{\rm Eucl}$ is a constant up to 
exponentially small corrections. In the regime 
$T>{ T_{1}}$ the Euclidean action
begins to decrease, 
 see Eq.~(\ref{otwet}). The temperature $T=T_{c}$ 
(see Eq.~(\ref{Tc}))
marks the second-order like 
transition from quantum to classical behavior.
}
\vskip0.3cm
\hskip-0.7cm

In conclusion, we have briefly summarized the Hall-creep
of pancake vortices relevant in clean high-temperature superconductors.


%

\end{document}